\documentclass[useAMS,usenatbib]{mnras}
\usepackage{epsfig,amssymb,amsmath,rotating,lineno,footmisc}
\usepackage{graphicx}
\newcommand\gs{{GS 1354-64}}
\newcommand\swift{{\it Swift}}
\newcommand\salt{{\it SALT}}
\newcommand\igl{{\it INTEGRAL}}
\newcommand\maxi{{\it MAXI}}
\newcommand\rxte{{\it RXTE}}
\newcommand\nus{{\it NuSTAR}}
\newcommand\xmm{{\it XMM-Newton}}
\title[X-ray and optical correlated analysis of GS 1354-64]{Simultaneous Optical/X-ray study of GS 1354-64 (=BW Cir) during hard outburst: evidence for optical cyclo-synchrotron emission from the hot accretion flow}

\author[Pahari et al.]{Mayukh Pahari$^{1,2}$\thanks{E-mail: mayukh@iucaa.in (MP)}, Poshak Gandhi$^{2}$, Philip A. Charles$^{2}$, Marissa M. Kotze$^{3,4}$,            
\newauthor
Diego Altamirano$^{2}$, Ranjeev Misra$^{1}$ \\
$^{1}$ Inter-University Centre for Astronomy and Astrophysics, Pune, 411007, India\\
$^{2}$ Department of Physics \& Astronomy, University of Southampton, Highfield, Southampton SO17 1BJ, UK \\
$^{3}$ South African Astronomical Observatory, PO Box 9, Observatory 7935, South Africa \\
$^{4}$ South African Large Telescope, PO Box 9, Observatory 7935, South Africa} 
\begin{document}

\pagerange{\pageref{firstpage}--\pageref{lastpage}} \pubyear{2016}

\maketitle

\label{firstpage}


\begin{abstract}

We present results from simultaneous optical (\salt{}) and X-ray (\swift{} and \igl{}) observations of GS 1354-64/BW Cir during the 2015 hard state outburst. During the rising phase, optical/X-ray time series show a strong anti-correlation with X-ray photons lagging optical. Optical and X-ray power spectra show quasi-periodic oscillations at a frequency of $\sim$ 18 mHz with a confidence level of at least 99\%. Simultaneous fitting of \swift{}/XRT and \igl{} spectra in the range 0.5$-$1000.0 keV shows non-thermal, power-law dominated ($>$ 90\%) spectra with a hard power-law index of 1.48 $\pm$ 0.03, inner disc temperature of 0.12 $\pm$ 0.01 keV and inner disc radius of $\sim$3000 km. All evidence is consistent with cyclo-synchrotron radiation in a non-thermal, hot electron cloud extending to $\sim$ 100 Schwarzschild radii being a major physical process for the origin of optical photons. At outburst peak about one month later, when the X-ray flux rises and the optical drops, the apparent features in the optical/X-ray correlation vanish and the optical auto correlation widens. Although $\sim$0.19 Hz QPO is observed from the X-ray power spectra, the optical variability is dominated by the broadband noise, and the inner disc temperature increases. These results support a change in the dominant optical emission source between outburst rise and peak, consistent with a weakening of hot flow as the disc moves in.
    
\end{abstract}
  

\section{Introduction}

In the field of accretion physics, substantial progress has been made in the last decade using correlated optical/X-ray studies which can connect inner accretion phenomena with outer accretion disc activity \citep{va94, de96,es00,ka01}. Recently, it has been realized that to fully understand the accretion and radiation mechanisms in the innermost part of the accretion disc (that emits mostly in X-rays), simultaneous, multi-wavelength observations are indispensable \citep[][ and references therein]{ut14, ru10}. One reason is the large variation in the time-scales involved in different physical processes that are responsible for emission other than X-rays. The observed range of time-scales can vary from a few tens of milliseconds to a few hundreds of seconds (e.g., \citet{ga16} and references therein). With new observatories such as AstroSat that facilitate high time resolution studies, the field of correlated optical/X-ray studies will become information rich, leading us to a clearer understanding of accretion structure and evolution. 

The origin of optical photons was long thought to be driven primarily by the reprocessing of hot, energetic, inner disc photons from the outer cold disc \citep{sh73}. In both cases, a rise in the X-ray flux will prompt a rise in the optical flux. Such phenomena lead to the detection of a strong, positive peak in the cross-correlation function (CCF) of the X-ray/Optical time series with some delay time, of the order of a few seconds to a few tens of seconds depending upon the size and orientation of the reprocessor. Recent simultaneous optical/X-ray observations from different black hole X-ray binaries (BHXBs) show that there exists a strong anti-correlation between optical/X-rays in the hard state with X-ray photon flares lagging optical photon deficits by a few seconds \citep{du08,ga08,ma03,ka01}. The observed anti-correlation is in sharp contrast with lagged positive correlation predicted by reprocessing. Interestingly, not only in BHXBs, but also a few neutron star X-ray binaries like Sco X-1, Cyg X-2 also show anti-correlation between optical/X-ray simultaneous time-series \citep{du11}. 

Sometimes, along with the anti-correlation, a strong, positive and narrow peak is observed in the CCF which complicates the interpretations of the origin of optical photons (e.g., \citet{ga08}). To interpret correlated optical/X-ray complex behaviour and the lag time scale which is of the order of fluctuation propagation time-scales \citep{ka01}, the X-ray photons are assumed to originate from the thermal Comptonization of the synchrotron radiation in an accretion disc corona \citep{mer00}. Alternatively magnetized outflows connecting both X-ray and optical emission locations are proposed to explain optical/X-ray correlated behaviour. However, both models fail to explain the `precognition dip', an anti-correlation prior to the strong lagged correlation, in the CCF. To explain this feature, a magnetic reservoir model has been proposed \citep{ma04}. This model assumes that an intense magnetic field can be generated using dynamo action in the disc, therefore magnetic flux tubes \citep{di99} of different scale heights sandwiching the disc, can store enormous amounts of energy to feed both jet and corona in a self-consistent manner \citep{ma04}. Both the observed optical/X-ray anti-correlation and the narrow peak can be explained by this model. Non-linear coupling of shots can further reproduce the so-called \lq rms--flux\rq\ relation \citep{ut05,ga09}. However, quasi-periodic oscillations (QPOs) with time-scales of milliseconds to tens of seconds in both optical \citep{mo83,du09,ga10} and UV \citep{hy03} bands are not inherent to the basic magnetic reservoir model. 

During the low/hard state, defined as an accretion state in X-ray binaries dominated by a flat powerlaw emission and a strong band limited noise in the Fourier power spectra \citep{re06}, the inner part of the truncated disc is usually filled with a hot, relativistic electron plasma \citep{yu14} along with intense magnetic fields \citep{es97}. The idea that such magnetized hot flow could be the birthplace of optical photons by means of electron-cyclotron emission, was introduced by \citet{fa82} and successfully explained some of the fast variability observed from GX 339-4 \citep{mo82}, 20-sec X-ray QPOs and the observed X-ray/optical anti-correlation in the CCF. Using reasonable assumptions, it was noted that the 20-sec QPO time-scale corresponds to the in-fall time in the hot flow region \citep{fa82}. One of the major predictions of the cyclo-synchrotron model is the detection of QPOs in both X-ray and optical with similar time-scales. According to the Lense-Thirring precession model \citep{st98,in09}, the generation of simultaneous optical and X-ray QPOs is possible within the hot flow due to the precession of the inner hot flow. \citet{ap84} shows that in the case of Cyg X-1, EUV photons produced by cyclotron emission can act as seed photons for high energy ($>$ 10 keV) Comptonization. In a series of works, \citet{ve11,ve13,ve15} have shown that cyclo-synchrotron optical photons undergoing Compton upscattering to X-rays in a hot flow can reproduce the observed optical/X-ray anti-correlation as well as QPOs. Additional component such as reprocessing can contribute a positive smeared optical response to the cross-correlation signal. Therefore, a hot flow model is particularly promising for objects which show anti-correlated optical/X-ray behaviour.

\begin{figure*}
\begin{center}
\includegraphics[scale=0.07]{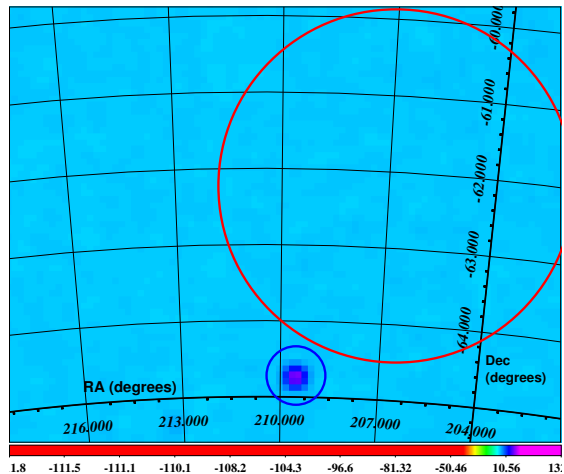} 
\caption{ 22.0-60.0 keV \igl{}/IBIS hard X-ray image of GS 1354-64 (circled in blue) constructed by superimposing all pointings. In the hard X-ray band, the source is significantly detected. The position of the historical \citep{ch67} soft X-ray transient Cen X$-$2 is shown by the red circle, and given the uncertainties involved is consistent with being the same source in spite of their different spectral properties (see \citet{ki90}). The circle size denotes the position uncertainty of Cen X$-$2 as quoted by \citet{ch67}. }
\label{image}
\end{center}
\end{figure*}                   

\gs{} (BW Cir) is a black hole X-ray transient that has had three X-ray outbursts over the past 28 years. During its 1997 outburst, \rxte{} detected moderate flux ($\sim$ 30-40 mCrab) but interestingly, throughout the outburst, the source remained in the hard state. Since \gs{} is close to the poorly located Cen X$-$2, a soft X-ray transient, discovered in 1967 with a peak X-ray flux of $\sim$ 8 Crab, it is likely they are the same source \citep{ki90}. Using the optical spectroscopy of GS 1354-64, \citet{ca09} presented dynamical evidence for a 7.9 $\pm$ 0.5 M$\odot$ BH. With P$_{orb}$ = 2.54 days \citep{ca09}, the donor is a G0-5 III star \citep{ca04} of mass 1.1 $\pm$ 0.1 M$\odot$ (from the rotational broadening of the companion's absorption spectrum). Interestingly, during quiescence, \gs{} shows strong optical variability \citep{ca09} with R filter falling by 4 magnitudes ($\sim$ 17 to $\sim$ 21) over a period of 14 years. \citet{ca09}, subsequently constrained the source distance and disc inclination angle to be $\ge$ 25 kpc and $\le$ 79$^{\circ}$ respectively. 

The outburst of GS 1354-64 in 2015 was first reported by \citet{mi15} when they detected a 0.5-10.0 keV X-ray flux 260 times higher than that in the quiescence. \swift{}/XRT X-ray spectrum on 10 June 2015 can be fitted with a power-law photon index of 1.5 $\pm$ 0.1 \citep{mil15}. Using the SOAR Optical Imager (SOI) at Cerro Pachon, Chile, \citet{co15} measured the g', r' and i' magnitudes to be $\sim$19.76, $\sim$18.49, $\sim$18.34 respectively on 10-11 June 2015. The detailed evolution of the 2015 outburst in GS 1354-64 at optical, UV and X-ray bands has been presented by \citet{ko16}. They find that the optical/UV emission is tightly correlated with the X-ray emission on time-scales of days.

In this work, we present simultaneous X-ray (\swift{}/XRT and \igl{}) and optical (\salt{}/BVIT) variability studies of GS 1354-64 during the 2015 X-ray outburst. We observe a strong anti-correlation between X-ray and optical on time-scales of $<$ 10 sec and harder X-rays show stronger anti-correlation with optical than soft X-rays. We perform simultaneous broadband spectral analysis in the energy range 0.5-1000.0 keV and show that all spectral and timing signatures strongly suggest that cyclo-synchrotron radiation in the presence of equipartition magnetic field is the major source of optical photons from the hot, optically-thin inner accretion flow during this hard state. We observed QPO like features at $\sim$18 mHz in the optical and X-ray power density spectra (PDS) with at least 99\% confidence level. With an increase in X-ray luminosity, the X-ray QPO frequency increases to $\sim$0.19 Hz while the optical PDS shows no QPO and no correlation/anti-correlation between X-ray and optical variability is observed. 

\section{Observations}

\subsection{{\it SALT}/BVIT analysis}

The optical observations of the source were taken on two nights $-$ 05 July 2015 and 08 August 2015 under clear sky conditions using the Berkley Visible Imaging Tube (BVIT) mounted on the South African Large Telescope (SALT) \citep{we12}. BVIT is a micro-channel plate photon counting detector with an active geometric area of 25 mm$^2$ and capable of tagging individual events with a precision of 25 ns \citep{we12,mc12}. Both observations of GS 1354-64 were binned to 10 ms time bins during the data reduction, using the IDL BVIT data-extraction pipeline \citep{mc12}. Details of observation are provided in Table \ref{obs}. During both observations, a comparison star was chosen with brightness nearly equal to that of the source. The first observation on 05 July 2015 was taken using no filter (white light) which made the instrument most sensitive around 500 nm with a strong response in the B, V bands extending through R up to 850 nm\footnote{\url {http://pysalt.salt.ac.za/data/scam/filters/}\label{note1}}. The neutral density (ND) was set to 0.5. The second observation was taken in R-band for which the response is centred on 650 nm\footref{note1} and the ND was set to 0.5. Barycentric corrections to the BVIT lightcurve are performed using the {\tt tcorr} tool available with the {\it ULTRACAM} data analysis package\footnote{\url {https://github.com/trmrsh/cpp-subs/blob/master/src/tcorr.cc}}.

\begin{figure*}
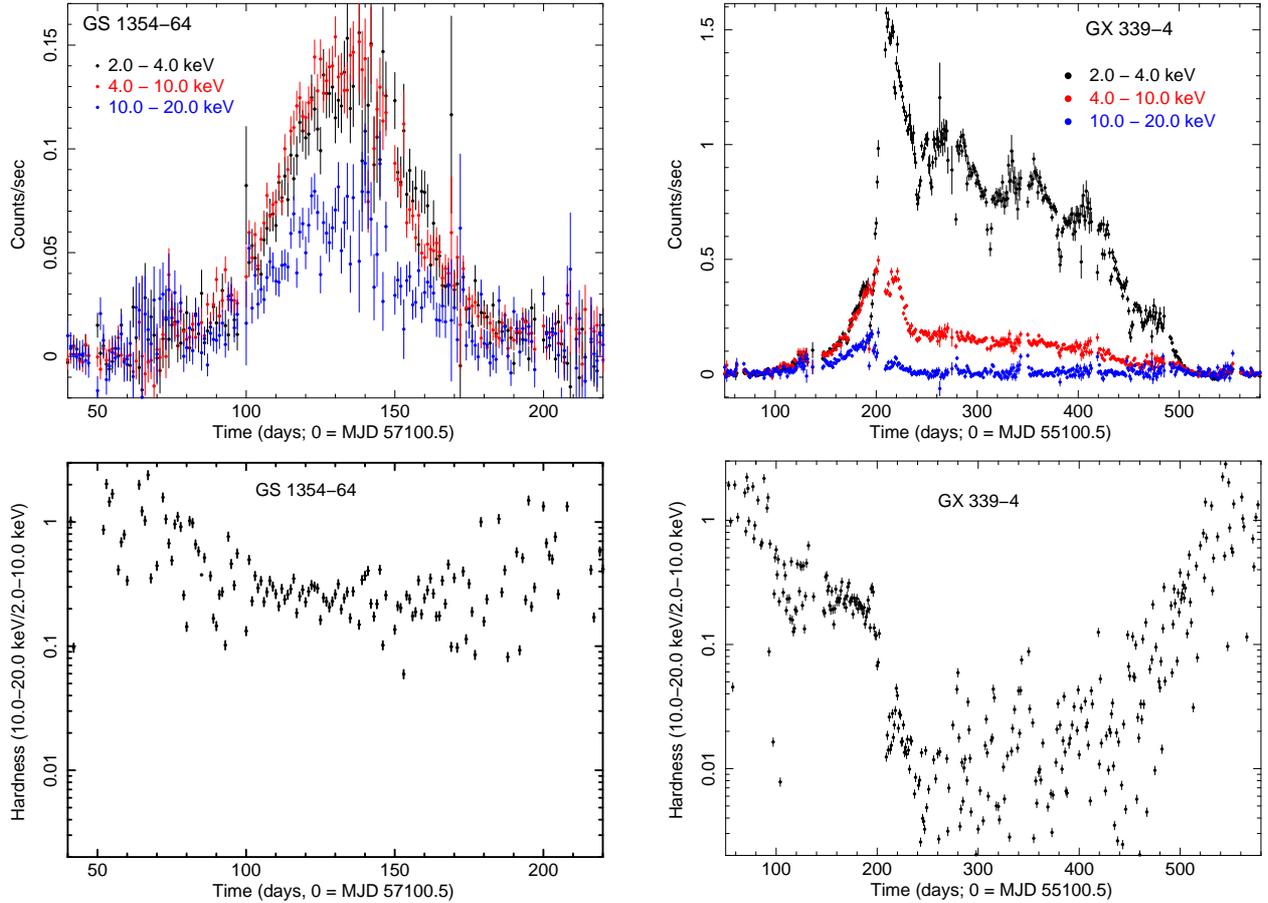

\begin{center}
\begin{tabular}{c|c}
\includegraphics[scale=0.33,angle=-90]{fig2a.ps} &
\includegraphics[scale=0.33,angle=-90]{fig2b.ps} \\
\includegraphics[scale=0.33,angle=-90]{fig2c.ps} &
\includegraphics[scale=0.33,angle=-90]{fig2d.ps} \\
\end{tabular}
\caption{The top left panel shows one-day averaged \maxi{}/GSC lightcurves, obtained in energy bands - 2.0-4.0 keV (black), 4.0-10.0 keV (red) and 10.0-20.0 keV (blue) during the 2015 outburst of GS 1354-64. To compare the hard outburst from GS 1354-64 with a typical X-ray outburst from the canonical X-ray binary GX 339-4, the MAXI lightcurves of GX 339-4 for its 2010 outburst are plotted in the top right panel in energy bands similar to that of GS 1354-64. The hardness ratio, (10.0-20. keV/2.0-10.0 keV) is shown as a function of time in the bottom left and bottom right panels for both outbursts from GS 1354-64 and GX 339-4 respectively. The drop in hardness by $\sim$2-3 orders of magnitude at the peak in GX 339-4 is clearly visible compared to the small drop in hardness at the peak in GS 1354-64. }
\label{hardness}
\end{center}
\end{figure*}

\subsection{{\it SWIFT}/XRT analysis}

{\it SWIFT}/XRT monitored the X-ray outburst of GS 1354-64 with an approximate cadence of 2 days. XRT provides simultaneous imaging and spectroscopy in the energy range 0.3$-$10.0 keV. To ensure pile-up free operations, all XRT observations were taken with Window Timing (WT) mode, and a time resolution of 17 msec. We follow standard procedures for extracting lightcurve, spectra and responses from XRT raw data.
Because of poor calibration and efficiency, counts in channels 0-29 ($<$ 0.3 keV) are ignored. All channels from 0.3-10.0 keV are binned such that a minimum of 30 counts per energy bin are available during spectral fitting. On 05 July 2015 and 08 August 2015 XRT observations were taken simultaneously with SALT and details are provided in Table \ref{obs}. Barycentric corrections to XRT photon arrival times are performed using the relationship (\swift{}/XRT team private communication):

\begin{multline}
T_1 = (t-T_{START})/86400 ~~sec \\
t_{COR} = T_{OFFSET} + (C_0 + C_1*T_1 + C_2*T_1^{2}) \times 10^{-6} ~~sec
\end{multline}

where

t is the time of interest, T$_{START}$ and T$_{OFFSET}$ are start time and offset time of the observation under consideration, t$_{COR}$ is the corrected photon arrival time and C$_0$, C$_1$ and C$_2$ are coefficients that can be obtained from the latest clock correction file {\tt swclockcor20041120v111.fits} updated after the observation and available on the HEASARC caldb website\footnote{\url{heasarc.gsfc.nasa.gov/FTP/caldb/data/swift/mis/bcf/clock}}. Because of calibration uncertainties below 0.5 keV \citep{pa15} and poor signal-to-noise above 8 keV, we use the spectra in the energy range 0.5-8.0 keV for model fitting. 

\subsection{{\it INTEGRAL} analysis}

{\it INTEGRAL} took 26 public pointing observations of GS 1354-64 between 05 July 2015 UTC 00:54:33 and 05 July 2015 UTC 22:11:34 during revolution no. 1560. The OMC, SPI, IBIS, JEMX1 and JEMX2 detectors of {\it INTEGRAL} were simultaneously on during each pointing. Observation details are provided in Table 1. All 26 archival data sets were processed and analysed using the INTEGRAL Offline Science Analysis (OSA; \citet{go03}) package v. 10.2, the {\tt Instrument Characteristics v 10.2} and the {\tt Reference Catalogue v. 40.0}. Following standard procedures, images are created by combining all science windows (each pointing lasts for 2-3 ks). Then they were cleaned and spectra extracted. In all four detectors - SPI, IBIS, JEM-X1 and JEM-X2 the source is detected with at least 6-sigma significance over the full energy range. To generate JEM-X1 and JEM-X2 spectra and corresponding responses, energy binning with up to 16 channels is used, giving an energy coverage of 3.0--35.0 keV. Since the low-energy threshold of IBIS/ISGRI on-board INTEGRAL has increased since launch, we ignore IBIS data below 22.0 keV. For SPI, we use the energy range of 25.0-1000.0 keV for spectral analysis.
 
\subsection{{\it NuSTAR} analysis}

\nus{} observed GS 1354-64 on several occasions during the 2015 outburst. Two observations on 11 July 2015 and 06 August 2015 were closest to the \salt{} observations presented here. \nus{} data are reduced using standard routines (\textsc{nupipeline} (\textsc{version 0.4.5}) and \textsc{nuproducts} (\textsc{version 0.3.0})) in the \nus{} Data Analysis Software \textsc{nustardas v1.6.0} included in $\textsc{heasoft v6.19}$ and the recent calibration database $\textsc{CALDB version 20161207}$. We use the background-subtracted, FPMA and FPMB combined lightcurve with 1 sec time resolution to derive the PDS in the 3-78 keV energy range.

\begin{table*}
 \centering
 \caption{X-ray and optical observation details of GS 1354-64 }
\begin{center}
\scalebox{0.95}{%
\begin{tabular}{ccccccc}
\hline 
Satellite/ & Instrument & Obs-ID  & Date & Start time  & Effective & average source  \\
Telescope & (mode) & & (dd-mm-yyyy) & (hh:mm:ss) & Exposure (sec) & count rate \\
\hline 
{\it Swift} & XRT (WT) & 00033811012 & 05-07-2015 & 18:30:28 & 965 & 14.9 $\pm$ 0.2 \\
SALT & BVIT (white light) & 20150705$\textunderscore$BWCir & 05-07-2015 & 18:18:11 & 2018 & 1014 $\pm$ 7 \\
{\it INTEGRAL} & JEMX1 & 12700040001 & 05-07-2015 & 01:06:19 & 66260 & 10.3 $\pm$ 0.6 \\
{\it INTEGRAL} & JEMX2 & 12700040001 & 05-07-2015 & 01:06:19 & 66160 & 10.4 $\pm$ 0.6 \\
{\it INTEGRAL} & IBIS & 12700040001 & 05-07-2015 & 01:06:19 & 52650 & 47.5 $\pm$ 0.2 \\
{\it INTEGRAL} & SPI & 12700040001 & 05-07-2015 & 01:06:19 & 53450 & 0.06 $\pm$ 0.01 \\
\hline
{\it Swift} & XRT (WT) & 00033811042 & 08-08-2015 & 18:09:30 & 507 & 28.7 $\pm$ 0.3 \\
SALT & BVIT (R-band) & 20150808$\textunderscore$BWCir & 08-08-2015 & 17:47:42 & 2236 & 382 $\pm$ 5 \\
\hline
\end{tabular}}
\end{center}
\label{obs}
\end{table*}

\subsection{Check for contamination}

The position of GS 1354-64 in the sky is often confused with the brightest soft X-ray transient Cen X-2 due to large uncertainties in the latter X-ray location. During its 1967 outburst, Cen X-2 showed a soft X-ray flux $\sim$50 times higher than that observed from GS 1354-64 and the spectra showed variations in power law indices between 1.15 and 2.8 \citep{co71,fr71}. On the other hand, GS 1354-64 is categorized as a hard X-ray transient which never reached the canonical high soft spectral states and state transitions. 
In Figure \ref{image}, a 22.0-60.0 keV \igl{}/IBIS cleaned image is shown combining all 26 exposures. This is the first hard X-ray image of GS 1354-64 observed with \igl{}. At the position of GS 1354-64, the source is detected with more than 10$\sigma$ significance (shown in blue circle) whereas within the red circle of Cen X-2 \citep{ch67}, no source is detected above the background limit. However, due to poor spatial resolution, the position uncertainties of the mean position of Cen X-2 are $\pm$ 2.5 degrees \citep{ch67}. 

Using \swift{}/XRT photon counting (PC) mode data, we are also able to extract the source image in the 0.3-10.0 keV energy range. We obtain a highly significant detection ($>$ 20 $\sigma$) of GS 1354-64 at the sky position identical to \igl{}/IBIS. However due to the strong, uncorrectable pile up, we cannot proceed further with PC mode data analysis. We find no other source in the field of view of \swift{}/XRT above the background level at 3$\sigma$ significance. 

\section{Timing analysis and results}

\subsection{Basic characterisation}

\begin{figure}
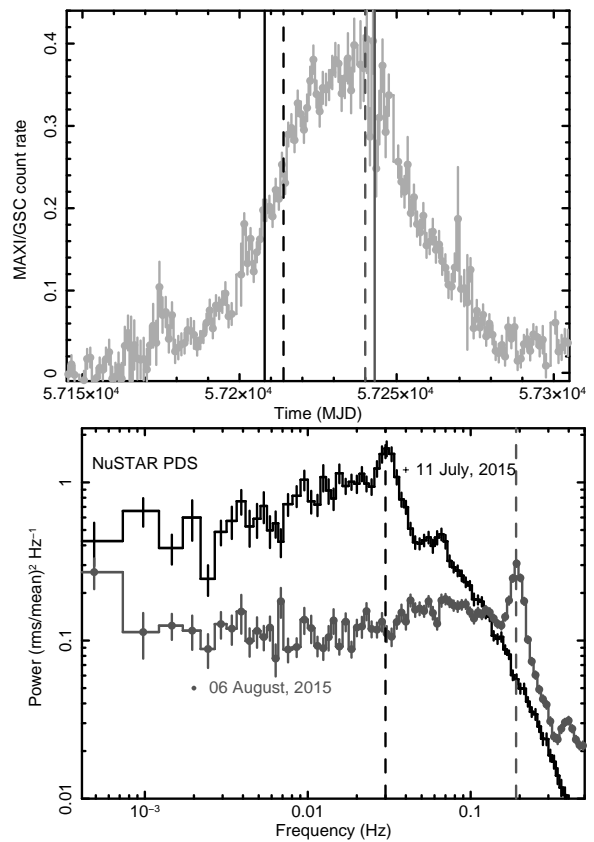

  \begin{center}
  \includegraphics[scale=0.31,angle=-90]{fig3a.ps} 
  \includegraphics[scale=0.31,angle=-90]{fig3b.ps} 
\caption{Top panel shows one day averaged {\it MAXI}/GSC lightcurve of GS 1354-64 during 2015 outburst in the energy range 2.0-20.0 keV. The times of two \salt{} observations, simultaneous with \swift{}, are marked by vertical straight lines and the times of two \nus{} observations are marked by vertical dotted lines. Bottom panel shows the rms-normalized and Poisson noise-subtracted power density spectra from two \nus{} observations on 11 July 2015 (marked by pluses) and 06 August 2015 (marked by filled circles) in the 3-78 keV energy range. Both PDS show the presence of strong QPO at $\sim$0.029 Hz and 0.19 Hz respectively. Top and bottom panels show that with the increase in X-ray flux QPO frequency increases. }
\label{maxi}
\end{center}
\end{figure}

In order to understand the nature of the outburst, we compare the \maxi{} one-day averaged lightcurve of the entire 2015 outburst of GS 1354-64 with the 2010 outburst from a canonical BHXB GX 339-4. In the top panels of Figure \ref{hardness}, the \maxi{} lightcurve of GS 1354-64 (top left) and GX 339-4 (top right) are shown in three different energy bands. From both panels it may be noted that 2.0-4.0 keV peak count rate in GX 339-4 is $\sim$3 and $\sim$8 times higher than that in 4.0-10.0 keV and 10.0-20.0 keV respectively while the same in GS 1354-64 is similar and $\sim$2 times respectively. This implies the behaviour of GS 1354-64 is much harder than GX 339-4. The spectral state evolution during both outbursts from GS 1354-64 and GX 339-4 is shown as the hardness ratio (ratio of count rate in 10.0-20.0 keV and 2.0-10.0 keV) plot in the bottom panels of Figure \ref{hardness}.  Transition from the hard to soft state is clearly observable in GX 339-4 as the hardness drops by $\sim$2-3 orders of magnitude at the peak count rate. But for GS 1354-64, the hardness ratio at the peak of outburst does not change dramatically beyond the hard state hardness level ($\sim$0.2).  
Therefore unlike GX 339-4, GS 1354-64 does not trace a canonical `q' shape in the hardness intensity diagram of BHXBs. This is a typical characteristic of the hard outbursts which are rare compared to that observed from GX 339-4. It is not clearly understood what is the driving parameter that causes variation in the nature of X-ray outbursts from different X-ray binaries, e.g., hard X-ray outbursts in some BHXBs (e.g., GS 1354-64, SWIFT J1753.5-0127 \citep{sh16}), while canonical soft X-ray outbursts in the majority of sources (e.g., GX 339-4 \citep{mi08}, XTE J1859+226 \citep{cas04}, XTE J1652-453 \citep{hi11} etc.) and a mix of hard and soft X-ray outbursts in a few other sources (e.g., H 1743-322 \citep{zh13} etc.). 

The times of the two \salt{} observations are shown by two vertical straight lines in the top panel of Figure \ref{maxi} where the 2.0-20.0 keV \maxi{} lightcurve of the entire 2015 outburst. The times of the two \nus{} observations are shown by dashed vertical lines. It may be noted that two observations were taken with \salt{} and \nus{} during two different X-ray flux states - the first during the rising phase of the outburst on 05 July 2015 (\salt{}) and 11 July 2015 (\nus{}) and the second at peak of the X-ray outburst on 08 August 2015 (\salt{}) and 06 August 2015 (\nus{}). If we compare the present outburst (also from Figure 1 in \citet{ko16}) with the 1997 outburst from \citet{br01}, the optical outburst peak occurs during the rising part of the X-ray outburst while a significant decline in the optical is observed during X-ray outburst peak. Therefore, an evolution in spectral and temporal properties is expected between both observations taken with a gap of $\sim$1 month. Figure 3 from \citet{st16} showed a systematic evolution of the powerlaw index and normalization during the entire outburst when they fitted the \swift{}/XRT spectra with an absorbed powerlaw model. Additional support for the evolution in parameter that describe QPO properties comes from the change in QPO frequencies in the PDS during both \nus{} observations which are shown in the bottom panel of Figure \ref{maxi}. At the low X-ray flux, a QPO is observed at 0.029 $\pm$ 0.004 Hz in the PDS during the first observation on 11 July 2015 while another QPO at the frequency of 0.19 $\pm$ 0.01 Hz is observed during the second observation on 06 August 2015 when the MAXI count rate in 2-20 keV energy range increases by a factor of $\sim$2.  

\subsection{Rising Phase Power Density Spectra}

\begin{figure*}
\begin{center}
\includegraphics[height=16cm,width=11cm,angle=-90]{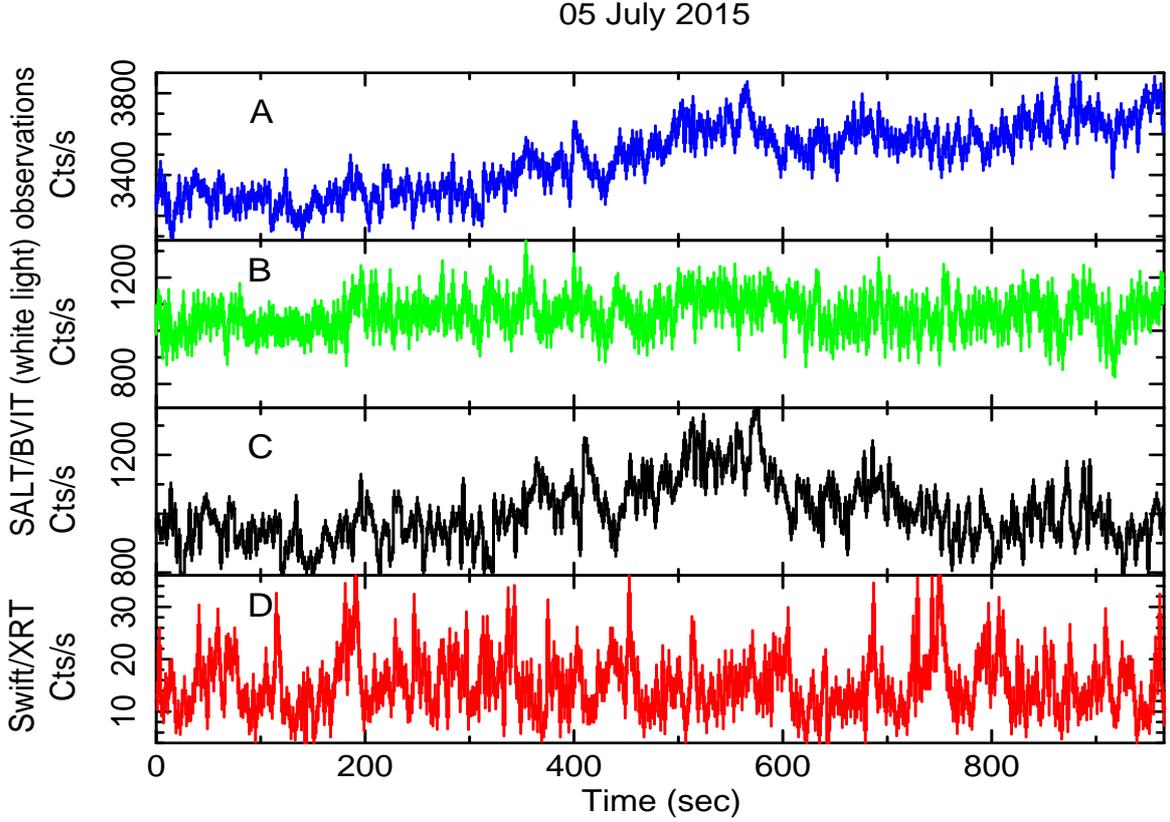} 
\caption{Strictly simultaneous lightcurves of GS 1354-64 as observed from \salt{}/BVIT (out of 2018 sec lightcurve) and \swift{}/XRT (out of 965 sec lightcurve) on 05 July 2015. Different panels show the raw, uncorrected {\it SALT}/BVIT optical lightcurve of GS 1354-64 (panel A), the background-corrected comparison star lightcurve (panel B), background \& atmospheric variation corrected source lightcurve (panel C) and background subtracted X-ray lightcurve from \swift{}/XRT (panel D). For clarity, all lightcurves are binned to 2 sec time resolution.}
\label{5jlight}
\end{center}
\end{figure*}

On 05 July 2015, the strictly simultaneous data obtained with \salt{}/BVIT and \swift{} are shown in Figure \ref{5jlight} using 2.0 sec binning. The uncorrected, raw source lightcurve, the background-corrected comparison star and background \& atmospheric variation corrected (divided by normalised comparison star) source lightcurve obtained from \salt{}/BVIT during the same observation are shown in panels A, B and C of Figure \ref{5jlight} respectively. The \swift{}/XRT lightcurve shows occasional short flares similar to that observed from the \salt{}/BVIT data. However, the flares are much stronger in terms of fractional X-ray flux, as compared to the optical flares which are relatively weaker. Fast variability of the order of few tens of seconds as well as slow variability of the order of few hundreds of seconds are observed in the optical lightcurve (panel C in Figure \ref{5jlight}). 
These long and short term variability components are absent in the comparison star lightcurve which implies that they are intrinsic to GS 1354-64. A close inspection of simultaneous X-ray and optical lightcurves reveals a hint of lagged anti-correlation (e.g., advancing the time axis of optical lightcurve by +0-20 sec, then short optical flares at $\sim$200 sec, $\sim$410 sec, $\sim$590 sec and $\sim$690 sec would correspond to dips in X-ray lightcurve at $\sim$205 sec, $\sim$420 sec, $\sim$610 sec and $\sim$710 sec respectively). 

To quantify the variability, we compute the power density spectra (PDS) of simultaneous X-rays (red) and optical (black) time-series on 05 July 2015 shown in the left panel of Figure \ref{pdsacf}. Both PDS are rms normalized and Poisson noise subtracted. A quasi periodic oscillation (QPO) like feature at 18 $\pm$ 1.3 mHz is observed in both PDS. The fractional rms amplitude (in percent) of X-ray and optical PDS are 23.8 $\pm$ 1.2 and 5.6 $\pm$ 0.8 respectively. We fit the PDS with two models : (1) power-law and (2) power-law + Lorentzian. For the optical, the change in $\chi^2$ = - 19 for change of 3 degrees of freedom {\it dof} when Lorentzian is included. An F-test between two models yields an F statistic value = 13.78 and the probability = 3.45 $\times$ 10$^{-6}$. In the case of the X-ray PDS, the change in $\chi^2$ = -29 for change of {\it dof} 5. The F-test yields F statistic value = 7.5 and the probability = 7.15 $\times$ 10$^{-5}$. In order to compute the confidence level on the detection of QPOs from X-ray and optical time-series, we follow the recipe for testing the significance of peaks in the periodogram of red noise data provided by \citet{va05}.  If the red noise can be fitted by a powerlaw continuum, then low significance peaks can be rejected accurately using his recipe. Using the powerlaw-fitted continuum PDS and computing confidence level, we find that the peak X-ray power in the PDS at the QPO frequency is higher than a 99.9\% confidence level while the peak optical power in the source PDS at the position of QPO frequency touches the 99\% confidence level. This is also consistent with the significance measurement method using fitted Lorentzian normalization and its error. The detection of the X-ray QPO with \swift{}/XRT at $\sim$18 mHz is also consistent with the $\sim$30 mHz QPO detected from the \nus{} observation taken 6 days later (on 11 July 2015; see bottom panel of Figure \ref{maxi}). When we extrapolate the trend observed from the QPO frequency vs time plot (as shown in Figure 1 from \citet{ko16}) to 05 July 2015, we find that the frequency predicted by the extrapolation matches exactly to the observed one.    

Near simultaneous detection of X-ray and optical QPOs have been reported earlier \citep{hy03}. Here, we observe strictly simultaneous potential QPOs in both X-rays and the optical at very similar frequency, implying similar characteristic timescales. An X-ray QPO at $\sim$ 18 mHz was also observed during the 1997 outburst of GS 1354-64 \citep{br01}.

\subsection{X-ray/optical correlation study during the rise}

To understand whether optical emission is the reprocessed X-ray from the outer accretion disc, we plot auto-correlation functions (ACFs; representing cross correlation of a time-series with itself) in the right panel of Figure \ref{pdsacf}. The ACFs in both optical and X-rays are distinctly similar to each other, and have a narrow core with full width at half maximum of 9.49 $\pm$ 1.16 sec. A simple reprocessing scenario over an extended accretion disc would be equivalent to a convolution of the X-ray (incident) lightcurve with a transfer function representing the response of an extended disc. In this case, we would expect the reprocessed (optical) ACF to be wider than the ACF of the incident (X-ray) emission. Given the large binary separation ($\sim$ 40 light-sec), we may expect reprocessing over the extended disc on timescales of order tens of seconds. The very close match between the optical and the X-ray ACFs would appear to rule this out. However, we note that complex interplay between separate out-of-phase components can simulate similar effects, even if reprocessing were present (e.g. \citet{ve11}). So the ACF analysis alone cannot rule out reprocessing completely.

We perform X-ray/optical cross correlation (cross-correlation coefficient as a function of time-delay between X-rays and optical lightcurve) which is shown in the left panel of Figure \ref{lag}. On the order of ten seconds, a strong anti-correlation is observed between X-rays and optical. A negative delay implies the X-ray band is delayed to the optical while anti-correlation implies out of phase variability between optical and X-ray. 

To compute the cross correlation, we used the tool {\tt pydcf}\footnote{\url{https://github.com/astronomerdamo/pydcf}} which is a python-based discrete cross correlation function which also works well with unevenly sampled data \citep{ed88,mc14}. We cross-checked our results with the {\tt crosscor} tool available in {\tt HEASOFT v. 6.19} and found that results from both tools match each other. The blue horizontal line in the left panel of Figure \ref{lag} marks the upper 99\% confidence level for significance of the cross correlation function based upon a two-tailed test. In this case, the approximate 99\% confidence interval (for which $\alpha$ = 0.01) is given by $\pm$ 2.58/$\sqrt N_s$ where {\it N$_s$} is the sample size of the time-series \citep{ch04}. Estimation of cross correlation coefficient outside the confidence interval is highly significant where the null hypothesis that the true cross correlation at a specified lag is zero must be rejected against the alternative hypothesis that the true coefficient is non-zero.

We note that the anti-correlation is the strongest feature to stand out in the cross-correlations, with no obvious, corresponding signature associated with a positive optical lag. This suggests that the anti-correlated component dominates the optical.

We also compute the cross correlation between the optical and relatively soft (0.3-2.0 keV) \& hard (2.0-8.0 keV) X-ray lightcurves from \swift{}/XRT separately, which are shown in the top and bottom right panels of Figure \ref{lag}, respectively. The soft X-ray band shows relatively weak anti-correlation with optical while the hard X-ray shows very strong anti-correlation (more significant than 99.7\% confidence level). This indicates that the process in the accretion flow where optical photons are generated is better coupled to hard X-ray generating region than soft X-ray.

\subsection{Checks on comparison star}

\begin{figure*}
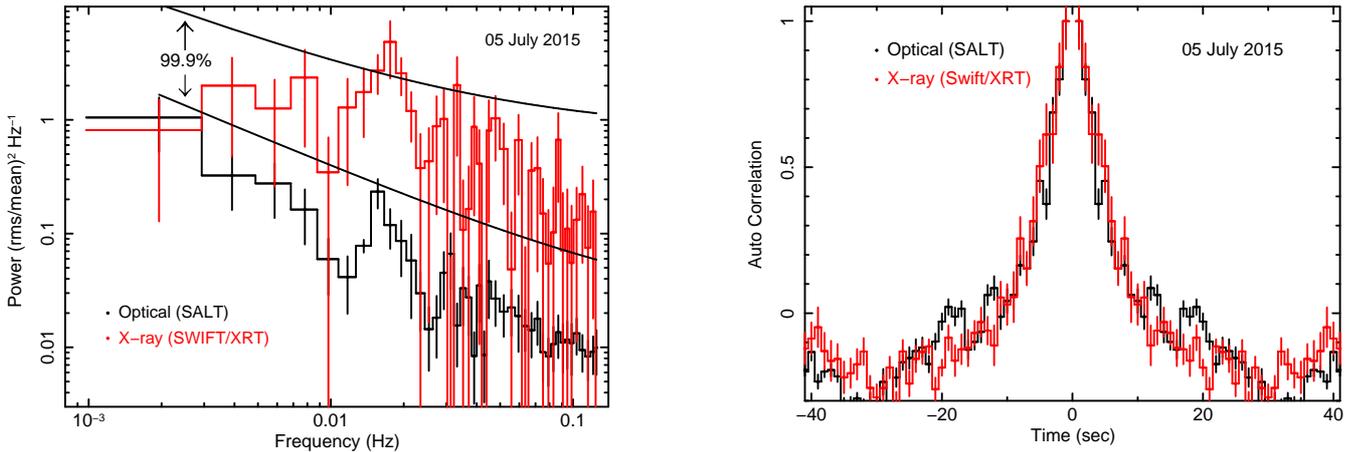

\begin{center}
\begin{tabular}{c|c}
\includegraphics[scale=0.335,angle=-90]{fig5a.ps} &
\includegraphics[scale=0.33,angle=-90]{fig5b.ps} \\
\end{tabular}
\caption{{\it Left panel}: Power density spectra (PDS) in the frequency range of 1 mHz to 100 mHz, obtained from simultaneous X-ray and optical time series as observed on 05 July 2015 are shown in red and black respectively. Both PDS are rms normalized and white-noise subtracted. QPO like features at $\sim$ 18 mHz are observed in both PDS with at least 99\% confidence. 99.9\% confidence levels are shown by black lines. {\it Right panel:} Auto-correlation function (ACF) of the X-ray (red) and optical (black) time series as observed on 05 July 2015. Both auto-correlation functions have similar width at least up to 30 sec which is an evidence against reprocessing from X-ray to optical.}
\label {pdsacf}
\end{center}
\end{figure*}

\begin{figure*}
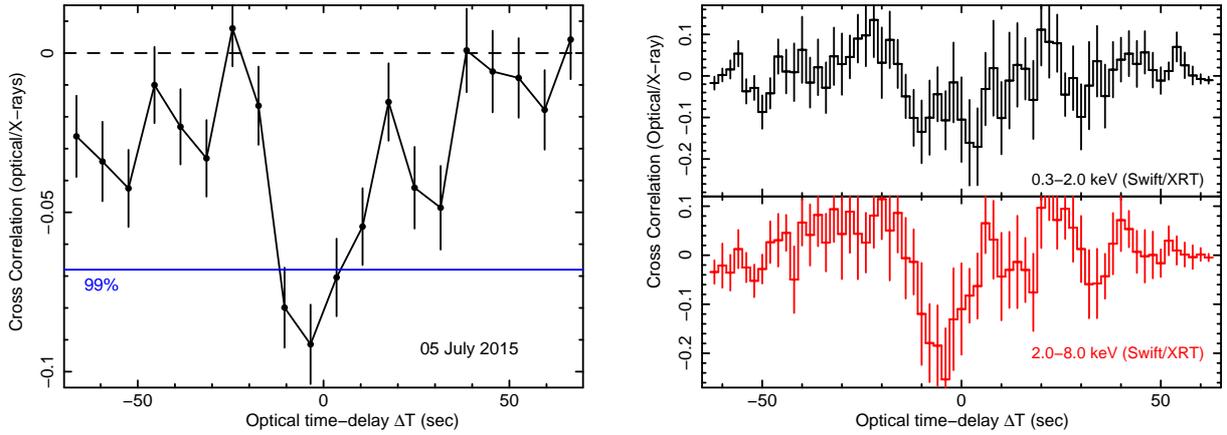

\begin{center}
 \begin{tabular}{c|c}
\includegraphics[scale=0.32,angle=-90]{fig6a.ps} &
\includegraphics[scale=0.32,angle=-90]{fig6b.ps} \\
\end{tabular}
\caption{Left panel shows the plot of cross correlation function (CCF) as a function of time delays between simultaneous X-ray (0.3-8.0 keV) and optical time series as observed on 05 July 2015. There is a strong anti-correlation observed between X-ray and optical lightcurves and the optical time series is delayed with respect to X-ray by $\le $ 10 sec. The blue horizontal line shows the 99\% confidence level of non-zero cross correlation coefficient. X-ray energy dependence of the CCF is shown on the right panel where the CCF is constructed between optical and soft X-ray energy bands (0.3-2.0 keV; top right panel) and between optical and hard X-ray energy bands (2.0-8.0 keV; bottom right panel). Stronger anti-correlation is observed with the hard X-ray energy band than the soft X-ray energy bands.}
\label{lag}
\end{center}
\end{figure*}

\begin{figure*}
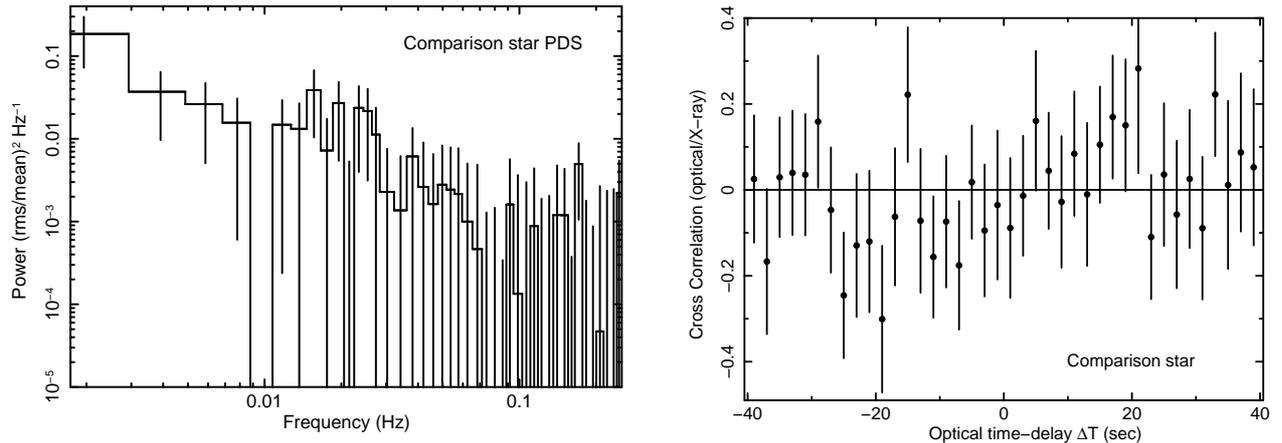

\begin{center}
 \begin{tabular}{c|c}
\includegraphics[scale=0.34,angle=-90]{fig7a.ps} &
\includegraphics[scale=0.32,angle=-90]{fig7b.ps} \\
\end{tabular}
\caption{The left panel shows the optical power density spectrum (PDS) of the comparison star in the frequency range 1 mHz to 100 mHz where no QPO like feature is observed. The right panel shows the plot of cross correlation function (CCF) as a function of time delays between simultaneous X-rays time-series (0.3-8.0 keV) of the source GS 1354-64 and the optical time series of the comparison star as observed on 05 July 2015. There is no significant anti-correlation observed between both lightcurves at any delay time confirming the anti-correlated lag observed in Figure \ref{lag} is due to the source only. }
\label{comp}
\end{center}
\end{figure*}

To confirm that observed features are not due to any instrumental, atmosphere or background artefacts, we compute the optical PDS of the background-corrected comparison star in the same frequency range as the source. In the comparison star PDS, shown in the left panel of Figure \ref{comp}, no QPO like features are observed with a confidence level higher than 90\% and the integrated rms power in the optical PDS is about one order of magnitude less than the optical rms power of GS 1354-64. This implies that the QPO like feature in the left panel of Figure \ref{pdsacf} is due to the source only. We also check the cross correlation between the simultaneous optical lightcurve of the comparison star and X-ray lightcurve of GS 1354-64 and found no significant anti-correlation which is shown in the right panel of Figure \ref{comp}. This implies that the lagged anti-correlation is due to the source.   

\subsection{X-ray high $-$ optical low state}

\begin{figure*}
\begin{center}
\includegraphics[height=16cm,width=11cm,angle=-90]{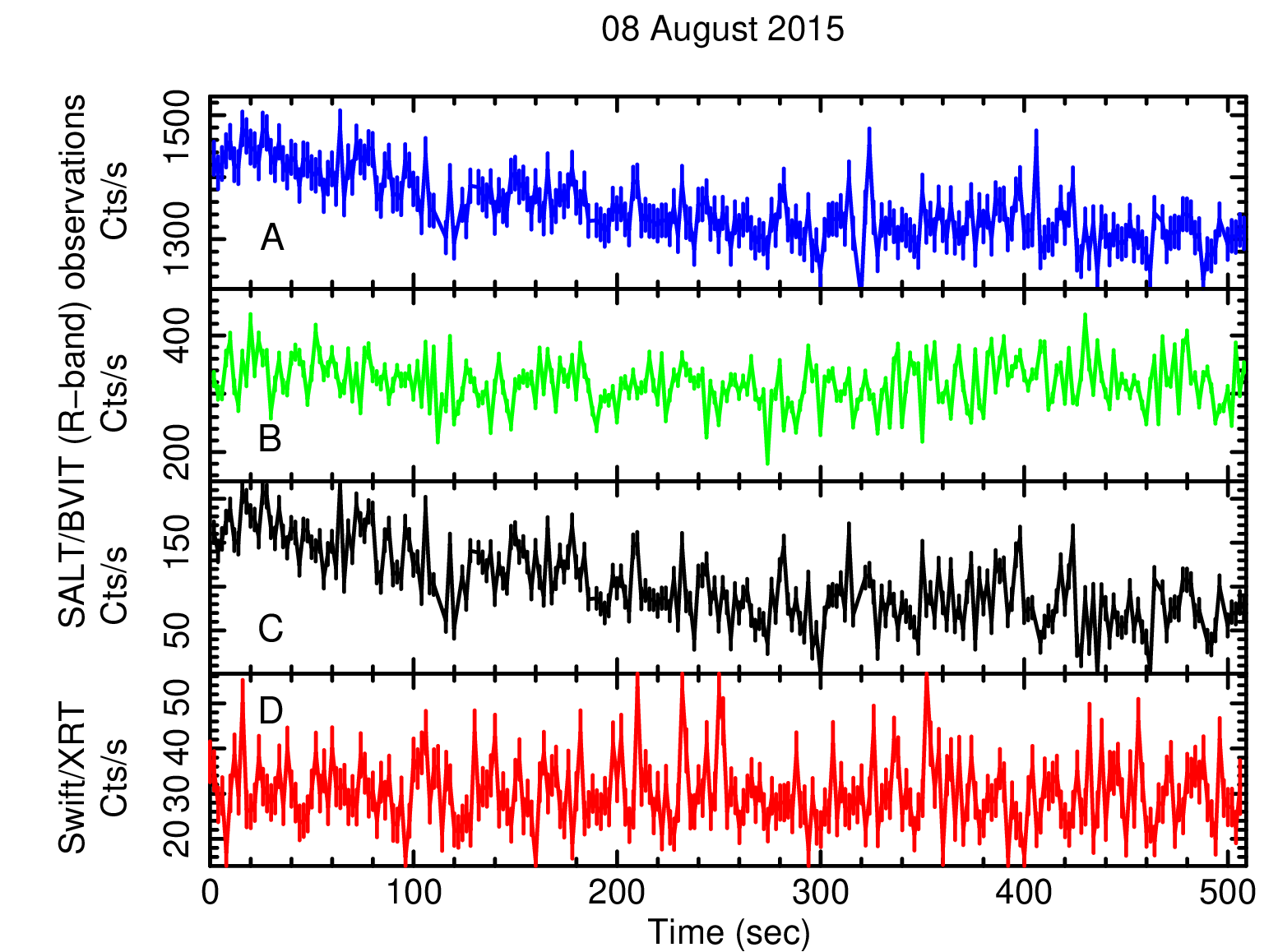} 
\caption{Strictly simultaneous lightcurves of GS 1354-64 as observed from \salt{}/BVIT (out of 2236 sec lightcurve) and \swift{}/XRT (out of 507 sec lightcurve) on 08 August 2015. Different panels show the raw, uncorrected {\it SALT}/BVIT optical lightcurve of GS 1354-64 (panel A), the background-corrected comparison star lightcurve (panel B), background \& atmospheric variation corrected source lightcurve (panel C) and background subtracted X-ray lightcurve from \swift{}/XRT (panel D). For clarity, all lightcurves are binned to 2 sec time resolution.}
\label{8auglight}
\end{center}
\end{figure*}

The second observation was taken on 08 August 2015 when the X-ray outburst reached the peak intensity (as observed from Figure \ref{maxi}). The optical lightcurves obtained from \salt{}/BVIT are shown in the top three panels (panel A, B and C) of Figure \ref{8auglight} while the background-subtracted, simultaneous \swift{}/XRT lightcurve in the energy range 0.3-8.0 keV is shown in the bottom panel (panel D).  

While the mean \swift{}/XRT count rate increases from $\sim$15 cts/s to $\sim$29 cts/s, the lightcurve presented in Figure 1 of \citet{ko16} indicates decrease in optical flux. Comparing BVIT counts appears to imply a decrease in optical flux between the first and second observation. But since our observations with BVIT on 05 July and 08 August were taken with different filters, we cannot directly compare optical fluxes from the BVIT data. Nevertheless a significant decrease in optical flux (by a factor of $\sim$3) on short time scales (in $\sim$500 sec) can be observed in the source lightcurve (Figure \ref{8auglight}) on 08 August. 

To determine the R magnitude during the second observation on 08 August 2015, we use SALTICAM images taken a few minutes before the BVIT observation. We use the finding chart provided by \citet{ca09} including field stars for performing relative photometry. For comparison we used nearby \lq Star D\rq\, as listed in Table 1 of \citet{ca09}, which has an R magnitude of 17.361. At the position of star D (R.A = 13:58:04.0 and Dec. = -64:44:02.0 (J2000)), the SALTICAM image shows the total count rate of 120 kcts/s within a circular region of 7 arcsec while at the position of GS 1354-64, the total count rate within 7 arcsec circle is 117 kcts/s. This gives an R magnitude of $\sim$17.384 for the optical counterpart of GS 1354-64 on 08 August 2015. This is very close to the R magnitude observed by \citet{ko16} when compared to their nearest observation on MJD 57245. The \swift{}/UVOT U filter flux densities from Figure 1 of \citet{ko16} are close to $\sim$0.055 mJy and $\sim$0.038 mJy during first and second observations respectively, consistent with the decrease in optical strength between two observations. Such optical/X-ray anti-correlated behaviour, although not widely studied, may be generic to the accretion process during hard X-ray outbursts.

\begin{figure*}
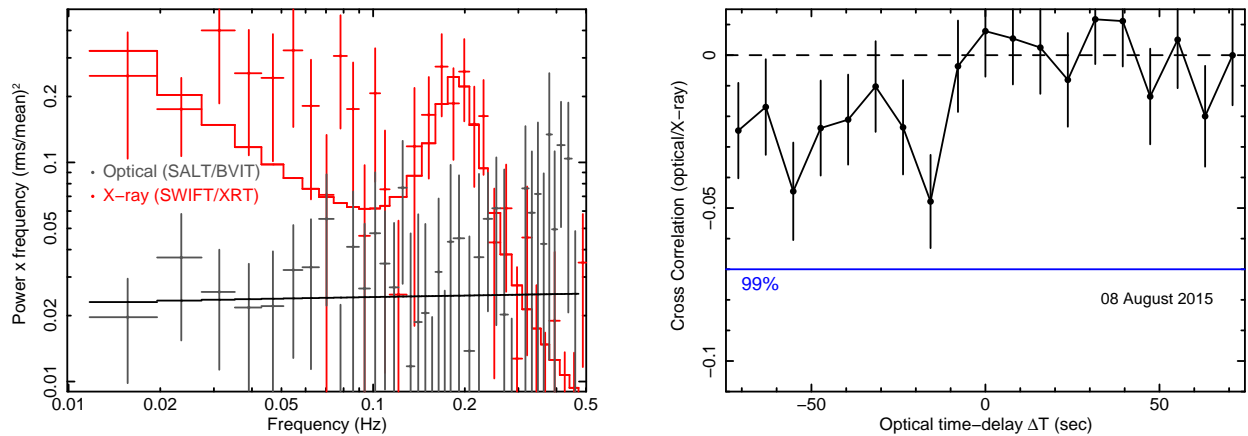

\begin{center}
 \begin{tabular}{c|c}
\includegraphics[scale=0.32,angle=-90]{fig9a.ps} &
\includegraphics[scale=0.32,angle=-90]{fig9b.ps} \\
\end{tabular}
\caption{{\it Left panel :} Power density spectra (PDS) in the frequency range of 1 mHz to 100 mHz, obtained from simultaneous X-ray and optical time series as observed on 08 August 2015 are shown in red and black respectively. Both PDS are rms normalized and white-noise subtracted. {\it Right panel:} Plot of cross correlation function (CCF) as a function of time delay between simultaneous X-rays (0.3-8.0 keV) and optical time series as observed on 08 August 2015 is shown. No significant anti-correlation or time delay is observed in the CCF.}
\label{8augpds}
\end{center}
\end{figure*}

We also have studied the PDS of X-ray (red) and optical (grey) time-series on 08 August 2015 which are shown in the left panel of Figure \ref{8augpds}. The X-ray PDS using \swift{}/XRT shows a strong QPO-like feature at $\sim$0.2 Hz. The QPO-like feature in the PDS is modelled with a Lorentzian while the continuum is modelled using a powerlaw. The combined model is shown by a red line. The model-fitted Lorentzian provides the QPO frequency of 0.189 $\pm$ 0.003 Hz, the QPO half-width at half-maxima of 0.028 $\pm$ 0.008 Hz and the QPO fractional rms of 14.3 $\pm$ 3.1 percent. A QPO at the similar frequency (0.19 $\pm$ 0.1 Hz) is also found in the \nus{} PDS observed two days before the second Swift/XRT observation presented here. Using \xmm{}/EPIC-pn observation of GS 1354-64 on 06 August 2015, \citet{st16} performed PDS analysis and found a QPO similar to that presented in our \swift{}/XRT PDS analysis. They observed (see Table 2) the QPO frequency of 0.192 $\pm$ 0.005 Hz,  the QPO half-width at half-maxima of 0.021 $\pm$ 0.007 Hz and the QPO fractional rms of 11.2 $\pm$ 1.3 percent. Additionally, they found a harmonic of this QPO at 0.399 $\pm$ 0.001 Hz with the fractional rms of 4.4 $\pm$ 0.8 percent and a low-frequency, broad Lorentzian noise component with the fractional rms of 23.3 $\pm$ 0.9 percent. However, due to the poor signal-to-noise in the \swift{}/XRT PDS on 08 August 2015, we are unable to detect additional components in the PDS. Therefore, QPO properties obtained from \xmm{} and \swift{}/XRT observations of GS 1354-64 with a gap of two days are qualitatively similar to each other.  The simultaneous optical PDS do not show any feature and can be fitted with a single powerlaw model which is shown by black horizontal line in the left panel of Figure \ref{8augpds}.

\begin{figure}
  \begin{center}
  \includegraphics[scale=0.31,angle=-90]{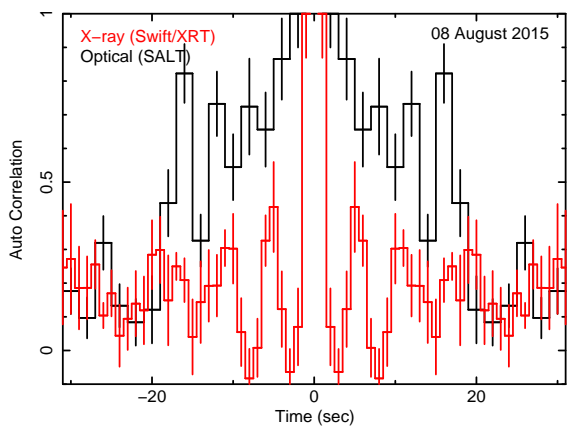} 
\caption{Auto-correlation function (ACF) of the X-ray (red) and optical (black) time series as observed on 08 August 2015. }
\label{acf8}
\end{center}
\end{figure}

In the right panel of Figure \ref{8augpds}, cross correlation between simultaneous X-ray and optical is shown as a function of time-delay. No features significant up to 90\% confidence are observed in the cross correlation pattern. Therefore, two important features $-$ (1) simultaneous QPOs from optical and X-ray variability and (2) strong anti-correlation between X-ray and optical lightcurves which are observed on 05 July 2015 are completely absent from X-ray and optical variability observed on 08 August 2015. This indicates that the increase in mass accretion rate from 05 July 2015 to 08 August 2015 causes QPOs to disappear in the frequency range between 1 mHz and 100 mHz. The integrated fractional rms (in percent) between 1 mHz and 100 mHz decreases from 21 $\pm$ 2 on 05 July to 8.8 $\pm$ 1.6 on 08 August. However, during both observations, the integrated fractional rms of the X-ray PDS is higher than that of the optical by the factors of $\sim$4-5. The comparison of X-ray and optical ACF is shown in Figure \ref{acf8}. At the time-scale $<$ 20 sec, The optical ACF is broader than X-ray ACF and notably it is broader than the optical ACF on 05 July 2015. Therefore, the optical state possibly changes between 05 July and 08 August. Broadening in optical ACF may be due to the reprocessing from large outer disc area possibly caused by the inward movement of the disc in response to the increased X-ray luminosity. The X-ray ACF shows wings with a quasi-periodic time-scale of $\sim$5 sec which is consistent with the observed QPO  at the frequency of 0.189 $\pm$ 0.003 Hz from the PDS.   

\section{Spectral analysis and results} 

\begin{figure}
\begin{center}
\includegraphics[scale=0.33,angle=-90, keepaspectratio]{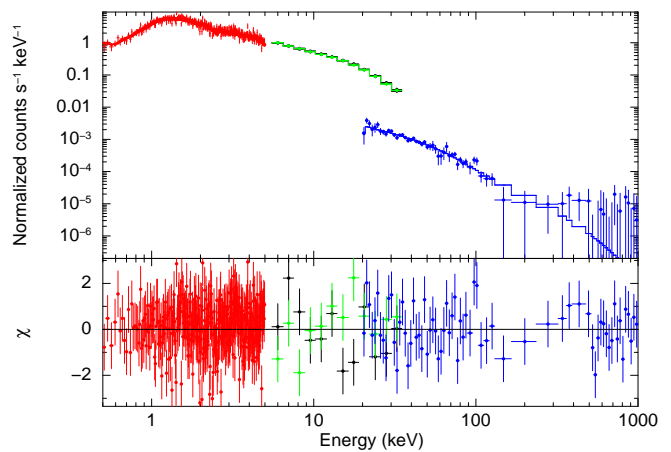}
\caption{Joint spectral fitting with simultaneous spectra obtained from \swift{}/XRT (red), \igl{}/JEMX1 (green), \igl{}/JEMX2 (black) and \igl{}/ISGRI (blue) on 05 July 2015 are shown in the top panel. For spectral modelling, disc blackbody {\tt diskbb} and thermal comptonization models {\tt nthcomp} are used. The bottom panel shows the residual of the fitting.}
\label{spec}
\end{center}
\end{figure}

To check the spectral-timing correlated evolution of the source, we perform the mean energy spectral analysis of simultaneous \swift{}/XRT and \igl{} in the energy range 0.5-1000 keV as observed on 05 July 2015. The aim of the joint spectral fitting is to understand the spectral nature of the underlying continuum during which optical and X-ray QPOs and the lagged anti-correlation are observed. Because of the hard nature of the outburst, the X-ray emission is expected to be dominated by the powerlaw-like component. Therefore, to fit the joint spectra, we use the \textsc{nthcomp} model in \textsc{XSPEC} which describes the hot flow emission as a thermal Comptonization of soft seed photons \citep{zd96}. The broadband spectra can be fitted with this model ($\chi^2$/dof = 317/285) above 2.0 keV. However, a strong residual below 2 keV is observed in the \swift{}/XRT spectrum which deteriorates the broadband fitting (0.5-1000 keV) to an unacceptable level ($\chi^2$/dof = 559/420). To account for this, we include the disc blackbody emission model \textsc{diskbb} which represents the emission from a cold accretion disc surrounding the hot flow. With the addition of \textsc{diskbb}, the fit improves significantly ($\chi^2$/dof = 463/418; change in $\chi^2$ = -96). To account for the Galactic neutral absorption, we use the \textsc{TBabs} model with column density allowed to vary during fitting. Cross calibration constants are used between various detectors. The best fit model yields an observed inner disc temperature of 0.12 $\pm$ 0.04 keV and measured inner disc radius of 3000 $\pm$ 500 km assuming a source distance of 25 kpc and inclination of 70$^{\circ}$ respectively. Both values represent a cold disc truncated at a large radius (translated to $\sim$100 R$_s$ assuming a black hole mass of 8.0 M$\odot$). The fitted \textsc{nthcomp} model component yields the photon power-law index of 1.48 $\pm$ 0.03 and Comptonizing, hot electron temperature of 90 $\pm$ 15 keV. These two parameters represent a hot inner flow with a very flat power-law like continuum. While fitting with the \textsc{nthcomp} model, we tied the Comptonization seed photon temperature to the inner disc temperature. The column density (N$_H$) is found to be 0.68 $\pm$ 0.08 $\times 10^{22} cm^{-2}$ which is consistent with the predicted Galactic absorption column density along the direction of GS 1354-64. Figure \ref{spec} shows the simultaneously fitted spectra from \swift{}/XRT, \igl{}/JEMXs and \igl{}/ISGRI in the top panel and the residuals of the fitted spectra are shown in the bottom panel.   

With the same best fit model, we fit the \swift{}/XRT spectra on 08 August 2015 in the energy range 0.5-10.0 keV since no simultaneous observation from any other instrument was present. The best fit yields an observed inner disc temperature to be 0.49 $\pm$ 0.06 keV and apparent inner disc radius to be 1300 $\pm$ 400 km. The power-law photon index in \textsc{nthcomp} model is 1.65 $\pm$ 0.05. Taking these at face value and comparing to those from July 05, we note that the cold, inner disc moves inward, resulting in a higher inner disc temperature and reducing the strength of the hot flow. This does not cause a state transition as during the entire outburst the source always remained in the luminous hard state \citep{st16} but may be enough to weaken any QPO from X-ray and optical bands below our detection limit. Consequently, the X-ray and optical that show opposite phases on 05 July 2015, lose this phase correlation on 08 August 2015 when the X-ray flux increases by a factor of $\sim$2.
   
\section{Discussion \& conclusion}
 
Using simultaneous X-ray (\swift{}/XRT and \igl{}) and optical (\salt{}/BVIT) data on two occasions, we have shown that during the rising phase of the X-ray outburst in GS 1354-64, the luminous hard state corresponds to the optical high state. Near X-ray peak, the optical has been shown to drop \citep{ko16}, suggesting a long-term anti-correlated connection between the bands.

\citet{st16} performed \swift{}/XRT spectral analysis of 65 observations covering nearly four months and starting from 10 June 2015 during the outburst in GS 1354-64. They modelled the spectra using an absorbed powerlaw. As observed from the Table 1 of \citet{st16}, the powerlaw photon index and the powerlaw normalization on 05 July 2015 vary in the range 1.42-1.51 and 0.154-0.175 respectively depending upon whether the absorption column density is kept free to vary or fixed at 8.6 $\times$ 10$^{21}$ cm$^{-2}$. These values are similar to the \swift{}/XRT and \igl{}/IBIS joint spectral analysis presented here. We obtain a powerlaw photon index of 1.48 $\pm$ 0.03 and powerlaw normalization of 0.165 $\pm$ 0.009 (in the unit of photons cm$^{-2}$ s$^{-1}$ keV$^{-1}$ at 1 keV) keeping the absorption column density as a free parameter. Using \swift{}/XRT spectra on 06 August 2015, they obtained the powerlaw index and the powerlaw normalization of 1.61 $\pm$ 0.01 and 0.396 $\pm$ 0.012 which are similar to the spectral parameters obtained from our analysis. Therefore, results from the spectral fitting presented here is consistent with that from \citet{st16}.
Using \nus{} spectra during low-flux (on 13 June 2015) and high-flux (11 July 2015) states, \citet{el16} detected an Fe emission line from both states and a Compton reflection hump from the high-flux hard state. They did not detect disc component in the spectra arguing that the disc emission is too low to be detected in the \nus{} band. This is consistent with the first \swift{}/XRT observation presented in this paper which was taken 6 days before the high-flux state \nus{} observation. The disc temperature with our \swift{}/XRT observation is 0.12 $\pm$ 0.01 keV. For such low disc temperature, if the disc spectrum is assumed to follow multi-temperature blackbody emission, it is unlikely that the disc blackbody would be detected with \nus{}. We find that a simple disc and thermal Comptonization model can fit the \swift{}/XRT spectra well with the photon powerlaw index of 1.65 $\pm$ 0.05 consistent with the powerlaw index 1.635(fixed) observed from the \nus{} spectral fitting.

During the rising phase of the X-ray outburst, a potential QPO is observed at $\sim$ 18 mHz in the X-ray and the optical (white light) PDS with at least 99\% confidence level which is also consistent with the QPO observed at $\sim$30 mHz from the \nus{} observation taken 6 days later. However, a similar QPO is not observed from the second optical (R-band) observation taken one month after the first observation. Earlier studies using optical observations of the BHXB GX 339-4 \citep{ga10} showed that a 50 mHz optical QPO is visible in the PDS with the ultraviolet (u'), green (g') and red (r') filters. Moreover the QPO in r' filter is the strongest (i.e., highest fractional rms) among three filters. QPOs were also detected from the simultaneous Optical/UV PDS from the BHXB XTE J1118+480 \citep{hy03}. Our \salt{} observations in July and in August were carried out using two different filters: white light and R, respectively. The above findings on other sources would argue that detection in one filter would favour the detection in the other filter also, all else being unchanged. Although nearly 50-60\% of the optical flux during our first observation (when the QPO is observed) is coming from outside the R-band, the non-detection of the QPO in the second observation with the R-band filter is intrinsic to the source properties and not due to the choice of filters. Since most of the variability power concentrates in the QPO which actually drives the observed correlation, the observed anti-correlation during the first observation and no correlation during the second observation are possibly due to the presence and absence of optical QPOs during both observations respectively. The X-ray spectral analysis shows that the X-ray state during the second observation is softer than the first observation. Such change in X-ray states may be attributed to the disappearance of the optical QPO. 

Best fit parameters from the broadband spectra in 0.5-1000 keV indicate the presence of a cold, truncated disc ($\sim$ 0.12 keV) with the hot, optically thin inner flow. The hot flow spectrum has a power-law photon index $\sim$1.48 and the electron temperature of $\sim$ 90 keV. Our energy spectral study shows that an optically thin hot electron corona can co-exist with the geometrically thin, cold disc. There is a model suggesting that the matter from the disc may be evaporated to feed the corona \citep{me00}. Depending on the mass flow rate, the accretion flow consists of a cool accretion disc which truncates at certain radius and evaporates into the inner hot Comptonizing plasma which fills up the inner part of the accretion flow \citep[][ and references therein]{do07}. Evidences of such truncation is observed from stellar mass BHXBs to low luminosity AGN like M87 \citep{re96}, M81, NGC 4579 \citep{de04} with the truncation radii ranging from 100 R$_{s}$ (R$_{s}$ is the Schwarzschild radius = 2GM/c$^2$) \citep{ga99,di99} to 1000 R$_{s}$. With increasing mass accretion rate, the inner edge of the cool disc usually moves in and the truncation radius decreases. If the QPOs are caused by the inner disc activity, then a higher QPO frequency is expected with the decrease in the inner disc radius (as reported by \citep{ko16}).

\subsection{Estimation of cyclo-synchrotron luminosity}

We have suggested that results from the variability analysis are consistent with a cyclo-synchrotron origin of optical photons. In order to confirm this, we estimate the flux of self absorbed cyclotron expected in the optical. Following the work of \citet{ta81}, 

\begin{equation}
F_{\rm cyclotron} = \frac{2\pi m_{*}^3 \nu_{c}^3 kT_{\rm e}}{3c^2}
\end{equation}      

where m$_*$ is the harmonic order of the cyclotron emission, $\nu_c$ is the cyclotron frequency of the emitted photons originating from the hot electron plasma and gyrating in the equipartition disc magnetic field B so that $\nu_c$ = eB/2$\pi$m$_e$; where e and m$_e$ are electron charge and electron rest mass respectively. The equipartition magnetic field at the inner disc for a moderately rotating black hole is estimated to be $\sim$10$^7$ Gauss assuming the turbulent viscosity in the disc is comparable to the magnetic viscosity (\citealt{sh73}). Recent accretion disc MHD simulations show that the magnetic field due to the MRI turbulence is of the order of 10$^6$ Gauss as required for powering jets (\citet{tc11} and references therein). However, the field strength of the disc magnetic field in the presence of a hot flow is not well known. Assuming a mass flow from the disc to the corona is comparable at the truncation radius, \citet{me02} show that B$_{\rm corona}^2$/8$\pi$ $\approx$ 10$^{-1.2}$ B$_{\rm disc}^2$/8$\pi$. Assuming B$_{\rm disc}$ $\sim$ 5 $\times$ 10$^6$ Gauss, we obtain B$_{\rm corona}$ $\sim$ 3.2 $\times$ 10$^6$ Gauss. Using this magnetic field, the cyclotron frequency we obtain $\nu_c$ = 4.6 $\times$ 10$^{12}$ Hz which corresponds to the emission in the far IR band. However, cyclotron emission is highly absorbed up to many higher order harmonics. For the cyclotron emission to appear in the optical, i.e., $\nu$ = 4.8 $\times$ 10$^{14}$ Hz, $\nu$/$\nu_c$ would be of the order of $\sim$ 104. For $\nu/\nu_c$ $\sim$ 100, it has been shown that the harmonic order of the cyclotron emission (m$_*$) would be $\sim$ 300 when kT$_e$/m$_e$c$^2$ $\sim$ 0.25 \citep{ta81,ta82}. With a decrease in the electron temperature, lower order harmonics appear at the similar emission coefficient. Therefore, we assume m$_*$ to be 250 since the model fitted electron temperature is $\sim$ 100 keV (i.e., kT$_e$/m$_e$c$^2$ $\sim$ 0.2). To compute the self-absorbed cyclotron flux (which we assume to be manifested as the optical flux at higher order harmonics) F$_{\rm opt}$, we use the following relationship \citep{ta81}:

\begin{multline}
 F_{\rm cyclotron} = F_{\rm opt} \simeq 4 \times 10^{4} \times (kT_{\rm e}/m_{\rm e}c^{2}) \times m_{*}^{3} \\ 
\times ( B_{\rm corona}/10^{4} \rm Gauss)^{3}  \rm ~ergs~s^{-1}~cm^{-2}
\end{multline}     

Using our earlier estimations of {\it B$_{\rm corona}$}, kT$_e$/m$_e$c$^2$ and m$_*$, we obtain {\it F$_{\rm opt}$} $\sim$ 2.5 $\times$ 10$^{17}$ ergs/s/cm$^2$. From the joint fitting of {\it Swift} and {\it INTEGRAL} X-ray spectra in the energy range 0.5$-$1000 keV, the inner disc radius is found to be 2900 $\pm$ 600 km (calculated from the normalization of the {\tt disbb} and the {\tt ezdiskbb} model assuming a colour correction factor of 1.7). If we assume the mass of the black hole to be 8.0 M$\odot$, then this radius translates to $\sim$ 100-130 R$_s$ (R$_s$ = Schwarzschild radius). If we assume that the inner part is filled with hot flow up to 100 R$_s$, then the optical luminosity would be L$_{\rm opt}$ $\simeq$ 3.31 $\times$ 10$^{35}$ ergs s$^{-1}$. 

From Figure 1 of \citet{ko16}, if we consider the {\it SMARTS} Bessel R magnitude of GS 1354-64 to be $\sim$17.18 on 05 July 2015, the corresponding optical flux will be $\sim$1.6$\times$ 10$^{-12}$ ergs s$^{-1}$ cm$^2$. Correcting for the reddening and assuming an approximate distance of $\sim$25 kpc, the optical luminosity would be $\sim$3 $\times$ 10$^{35}$ ergs s$^{-1}$ which is very close to our estimation from theoretical considerations as well as fitted spectral parameters. From the {\it Swift}/XRT and {\it INTEGRAL} joint spectral fitting, 0.1$-$1000 keV unabsorbed X-ray flux is found to be 2.51 $\times$ 10$^{-8}$ ergs s$^{-1}$ cm$^2$. If we assume the distance to the source to be 25 kpc, the X-ray luminosity would be L$_{\rm x-ray}$ $\simeq$ 14.14 $\times$ 10$^{37}$ ergs s$^{-1}$. Therefore, L$_{\rm opt}$/L$_{\rm x-ray}$ $\simeq$ 0.21\% and the origin of optical photons is highly consistent with the cyclo-synchrotron process. Being the ratio of two luminosities, it is independent of the distance to the source. 

\subsection{Comparison with 1997 outburst}

During the 1997 hard state outburst of GS 1354-64 \citep{br01}, when \rxte{}/ASM (2-12 keV) and {\it CGRO}/BATSE (20-2000 keV) fluxes reached the peak X-ray flux of the outburst, the B, R and V band fluxes decreased from their respective peak values (from Figure 1 of \citet{br01}). Interestingly, this also matches with our SALT observations and those taken by \swift{}/UVOT \citep{ko16} during the latest outburst. In addition, the X-ray lightcurve profile of the current outburst also matches with that during 1997 (Figure 1 \citet{br01}). The evolution of accretion mechanism during a outburst decides the shape of the profile and it has been observed that different outburst profiles from the same source are usually different \citep{re06}. However, the similarity in the outburst profiles in GS 1354-64 at different epochs indicates that the accretion mechanism during different hard X-ray outbursts may be very similar. It is also clear that radio jets are observed during the rising phase of the 1997 X-ray outburst and falls rapidly afterwards. If we assume that 1997 and 2015 outbursts in this source follow similar accretion-ejection evolutionary paths, then we expect strong jet activity during our first optical observation. Therefore, the formation of jet base (probably a large spherical corona) which may provide strong non-thermal Comptonization of seed photons (either optical/UV photons generated inside the hot flow or photons from the cold disc surrounding the hot inner flow) is possible during our simultaneous \swift{}, \salt{} and \igl{} observations. Evidence for such jet-base/corona formation may comes from the very flat power-law index ($\sim$ 1.48) and very hot coronal temperature ($\sim$ 100 keV) obtained from the joint \swift{} and \igl{} spectral fitting, but this needs further investigation. 

The spectral state during simultaneous \swift{}, \salt{} and \igl{} observations can be described as the luminous hard state \citep{yu04} in which the nature of accretion is not well-understood. Assuming the compact object mass of GS 1354-64 to be 8 M$\odot$, the Eddington luminosity would be 10.1 $\times$ 10$^{38}$ ergs/s. Therefore, the observed X-ray luminosity, obtained from spectral fitting is $\sim$ 14\% of the Eddington luminosity (L$_{\rm Edd}$). 

\subsection{On the origin of X-ray/optical QPOs}

The $\sim$55 sec QPOs observed from optical and X-ray lightcurves are expected from the variation in the size of the hot inner flow \citep{fa82}. In the present work, if we assume the size of the hot inner flow is same as the disc truncation radius which is $\sim$ 3000 km, then following the argument of \citet{fa82}, the in-fall time-scale can be calculated as high as 50-60 sec provided the hot flow filling factor could be very close to unity. It is interesting to note that the excess in the optical and X-ray power density spectra at similar frequencies ($\sim$ 8 mHz) has also been reported by \citet{ve15} in SWIFT J1753.5$-$0127 and they interpreted the simultaneous X-ray and optical excess as the Lense$-$Thirring precession of the inner hot flow \citep{st98,in09}. Precession of the hot inner flow may play an important role here because we observe X-ray and optical simultaneous QPOs at the time-scale of $\sim$ 55 sec during a bright hard state. If the optical QPO originates within the hot flow by means of the cyclo-synchrotron process and the observed QPO time-scale is identical to typical Lense-Thirring time-scales, then precession is a natural explanation of the observed QPOs.

Hard X-rays are found to have a stronger anti-correlation with the optical than soft X-rays. This is naturally expected in the synchrotron self-Compton model of \citet{ve11} with the emitted spectrum pivoting at intermediate frequencies and varying strongly at the extreme ends of the optical and the X-ray regimes. It could also be related to stratification of physical conditions across the compact hot flow, with the optical possibly originating in the outer, colder region and the hard X-rays from the inner (hottest) regions. Alternatively, an independent disc contribution to the soft X-ray regime could also be responsible for weakening the soft X-ray/optical connection. Therefore, the origin of the soft photon could be the cold outer disc, rather than some mechanism associated with the precessing hot flow. However in an alternate scenario, weakening of anti-correlation in softer X-ray band has been explained by the cyclo-synchrotron self-Compton-disc reprocessing model by \citet{ve11}. 
        
Between the first and second optical observations while the X-ray outburst reaches the peak intensity, the QPO from the X-ray lightcurve shifted to higher frequency (nearly by a factor of 10) and the QPO in the optical band disappeared from the PDS. X-ray spectral analysis shows that the inner disc temperature increase by a factor of $\sim$4 between first and second observations. Therefore, the enhanced inner disc activity as it moves inward may inject instabilities that weaken such hot flow precession and may cause the disappearance of the optical QPO. Disruption of hot flow precession can also be caused by the radiation pressure instability since the luminosity at the peak of the X-ray outburst reached $\sim$ 0.4 $L_{\rm Edd}$. Exploring the origin of QPOs further is beyond the scope of the present work.

Missions like {\it AstroSat} which has optical/UV and X-ray detectors with the capability of event recording with high time resolution will be able to measure optical and X-ray variability simultaneously from different BHXBs. This would be immensely important for deeper understanding of accretion mechanism during the luminous hard state.     

 \section*{Acknowledgements}

MP acknowledge that a major part of this work is done under UGC-UKIERI thematic partnership grant UGC 2014-15/02 at the School of Physics \& Astronomy, University of Southampton, UK. MP is thankful to the SWIFT and INTEGRAL team members for arranging X-ray observations simultaneous to the SALT telescope observation on short notice. BVIT observations were obtained with the Southern African Large Telescope (SALT) under program 2014-1-MLT-003 (PI: M.M. Kotze). MP also thanks SAAO team for making multiple efforts for the observation due to bad weather. PG acknowledges funding from STFC (ST/J003697/2) and thanks T.R. Marsh for the public ULTRACAM software. P.G. thanks STFC for travel support to foster international collaboration. DA acknowledges support from the RS. This research has made use of data obtained through the High Energy Astrophysics Science Archive Research Center online service, provided by the NASA/Goddard Space Flight Center.

\bsp

\label{lastpage}

\end{document}